\documentclass[epj]{svjour}
\usepackage{graphics}

\newcommand{\AmS}{{\protect\the\textfont2
  A\kern-.1667em\lower.5ex\hbox{M}\kern-.125emS}}

\begin{document}

\title{Hadronic structure from two photon collisions}

\author{M.~R. Pennington}
\institute{Institute for Particle Physics Phenomenology,\\ Physics Department, Durham University, 
        Durham DH1 3LE, U.K.}
\date{Received: date / Revised version: date}   
       

\abstract{
We briefly describe what a two photon capability with KLOE2 can measure and discuss what this will teach us about the world of light hadrons.}


\maketitle

\section{Introduction}

Two photon physics has been explored by nearly all $e^+e^-$ colliders, ever since it was realised that significant production rates could be achieved~\cite{brodsky}. DA$\Phi$NE is to date the exception. Only now with the KLOE2 upgrade is there the possibility of installing a two photon capability. Accessing $\gamma\gamma$ centre-of-mass energies up to 1.1 GeV with such a detector will bring the exciting prospect 
that all of  $\pi^+\pi^-$, $\pi^0\pi^0$, $\pi^o\eta$, $K^+K^-$ and ${\overline{K^0}}K^0$ final states can be studied, and resonances in these channels detected.

The
two photon production of hadronic resonances is often advertised as one of the clearest ways of revealing their composition~\cite{klempt,mp-menu,barnes,achasov,hanhart,sa,menn,achasov-belle,giacosa}. For instance, the nature of the isoscalar scalars  seen in $\pi\pi$ scattering below 1.6 GeV, the $f_0(600)$ or $\sigma$, $f_0(980)$, $f_0(1370)$ and $f_0(1510)$ mesons,  remains an enigma \cite{klempt,mp-menu}. While models abound in which some are ${\overline q}q$, some ${\overline{qq}}qq$, sometimes one is a ${\overline K}K$-molecule, and one a glueball~\cite{klempt}, definitive statements are few and far between.
Their two photon couplings will help.

\begin{figure}[h]
\begin{center}
\resizebox{0.72\columnwidth}{!}{
\includegraphics{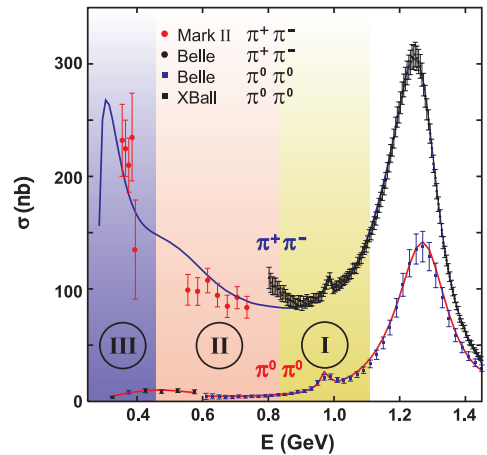}
}
\caption{Cross-section results for 
$\,\gamma\gamma\to\pi^+\pi^-\,$ from Mark~II~\cite{boyer} (below 800 MeV) and Belle~\cite{abe} above, integrated over 
$|\cos \theta^*|\,\le\,0.6$, and for $\gamma\gamma\to\pi^0\pi^0$ from Crystal
Ball~\cite{cb88} below 600 MeV and Belle~\cite{belle00} above, integrated over
$|\cos \theta^*|\,\le\,0.8$. $E$ is the $\gamma\gamma$ c.m. energy. The curves are from the (as yet) unpublished Amplitude Analysis that includes the data of Ref.~\cite{belle00}. The three shaded bands delineate the energy regions where KLOE2 can make a contribution discussed in the text.}
\label{fig:xsects}
\end{center}
\vspace{-8mm}
\end{figure}

The ability of photons to probe such structure
naturally depends on the photon wavelength. This is readily illustrated by looking at the
integrated cross-sections from Mark II~\cite{boyer}, Crystal Ball~\cite{cb88} and Belle~\cite{abe,belle00} shown in Fig.~1 for $\gamma\gamma\to\pi^+\pi^-$, $\pi^0\pi^0$. Let us think about these processes from the crossed channel viewpoint in which the photon scatters off a pion. At low energies the photon has long wavelength, and so sees the whole hadron and couples to its electric charge. Thus the photon sees the charged pions. The $\pi^+\pi^-$ cross-section is large: how large is a measure of the charge on the pion. In contrast, the neutral pion cross-section is small. However, as the energy increases the photon wavelength shortens and recognises that the pions, whether charged or neutral, are made of the same charged constituents, namely quarks, and 
causes these to resonate, as illustrated in Fig.~2. Thus both channels reveal the well-known ${\overline q}q$ tensor meson, the $f_2(1270)$, seen as a peak in 
Fig.~1,
with its production cross-section related to the average charge squared of its
constituents.

\begin{figure}[h]
\begin{center}
\resizebox{0.45\columnwidth}{!}{
\includegraphics{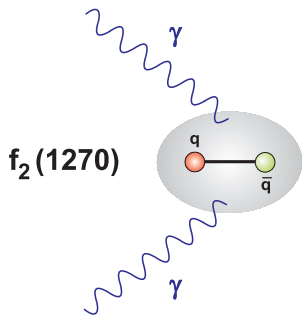}
}
\caption{Illustration of two photons coupling to a relatively long-lived
${\overline q}q$ state. The photons predominantly couples to the electric charge of the constituents.}
\label{fig:f_2}
\end{center}
\end{figure}

However, at lower energies, 500-1000 MeV, the photon wavelength is longer. States like the $\sigma/f_0(600)$ are so short-lived that they very rapidly disintegrate into two pions and when these are charged, the photons couple to these, Fig.~3. The intrinsic make-up of the state, whether ${\overline q}q$, ${\overline{qq}}qq$ or glueball, is  obscured by the large coupling to the pions to which the $\sigma$ decays. Data reveal a similar situation applies to the heavier, and seemingly much narrower, $f_0(980)$. This state has equally large hadronic couplings and is only narrow because it sits just below ${\overline K}K$ threshold, to which it strongly couples.
Experimental results discussed below suggest the two photons largely see its meson decay products too, regardless of whether the $f_0(980)$ is intrinsically a ${\overline K}K$ molecule or not~\cite{barnes,achasov,hanhart,achasov-belle,giacosa}.
\begin{figure}[bh]
\vspace{1mm}
\begin{center}
\resizebox{0.45\columnwidth}{!}{
\includegraphics{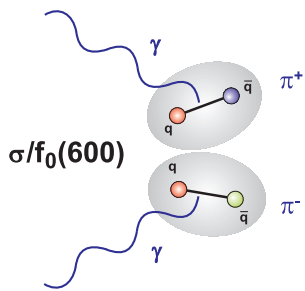}
}
\caption{Illustration of two photons coupling to a state that quickly decays to $\pi\pi$. In the case of the $\sigma/f_0(600)$, the photons coupling to the final state dominates over any coupling to the {\it intrinsic} make-up of the state,
regardless of its composition.
}
\label{fig:sigma}
\end{center}
\end{figure}

\section{Amplitude Analysis}

In the future, strong coupling QCD will eventually predict the two photon couplings of these states according to models of their composition. While theorists work on this, what we have to do
 is to determine the resonance couplings from experiment.
To extract these reliably requires a partial wave separation.
In principle one needs data with full angular coverage with polarised photon beams. But we have no polarisation information and in the two photon centre-of-mass frame at most 80\% angular acceptance, less in the case of $\pi^+\pi^-$ because of the difficulty of separating these from the scattered $e^+$ and $e^-$. Thus even for the large $f_2(1270)$ signal seen so prominently in the integrated cross-sections of Fig.~1 determining its two photon width is not so easy. One must separate the $\pi\pi$ amplitude into components with definite spin, helicity and isospin.

 The $\pi^+\pi^-$ cross-section near threshold is dominated by the one pion exchange Born amplitude, producing the enhancement seen in Fig.~1. Being controlled by $I=1$ exchange in the crossed channels means that at low energies $\pi\pi$ production in $I=0$ and $I=2$ must be comparable in all partial waves.
Thus data on one charged final state cannot be meaningfully analysed on their own.
 
The era of high luminosity $e^+e^-$ colliders with their intense programme of study of  heavy flavour decays has, as a by-product, yielded two photon data of unprecedented statistics. The Belle collaboration~\cite{mori,abe} have published results on $\gamma\gamma\to\pi^+\pi^-$ in 5 MeV bins above 800 MeV. These show a very clear peak for the $f_0(980)$, Fig.~1. Belle~\cite{abe}, analysing just their integrated cross-section, find its radiative width to be $205 ^{+95+147}_{-83-117}$ eV.
 The large errors reflect the many ways of drawing a background  whether in the $I=0$ $S$-wave where the resonance appears or in the other partial waves: remember that without full angular coverage the partial waves are not orthogonal, and so interferences occur. Despite these uncertainties,
a number of theoretical predictions have now honed in on $0.2-0.3$ keV for the radiative width of the $f_0(980)$, whether it is a ${\overline K}K$ molecule or a $\overline{qq}qq$ state~\cite{hanhart,sa,giacosa}. 

The only way to make sense of the real uncertainties is to perform an Amplitude Analysis.
A key role is played by the general $S$-matrix properties of analyticity, unitarity and crossing symmetry.  When these are combined with the low energy theorem for Compton scattering,
these anchor the partial wave
  amplitudes close to $\pi\pi$ threshold~\cite{morgam} as described in Refs.~\cite{mpdaphne,belle-mp} and so help to make up for the lack of full angular coverage 
in experiments. Crucially, unitarity imposes a connection between the $\gamma\gamma\to\pi\pi$ partial wave amplitudes and the behaviour of  hadronic processes with $\pi\pi$ final states. Below  1~GeV the unitarity sum is saturated by the $\pi\pi$ intermediate state, while above the ${\overline K}K$ channel is critically important. Beyond 1.4-1.5 GeV multipion processes start to contribute as $\rho\rho$ threshold is passed. Little is known about the $\pi\pi\to\rho\rho$ channel in each partial wave. Consequently, we restrict attention to the region below 1.4 GeV, where $\pi\pi$ and ${\overline K}K$ intermediate states dominate.
The  hadronic scattering amplitudes for $\pi\pi\to\pi\pi$ and ${\overline KK}\to\pi\pi$ are known and so enable the unitarity constraint to be realised in practice and in turn allow an Amplitude Analysis to be undertaken.

Such an analysis has been performed~\cite{belle-mp} incorporating all the world data and its key angular information~\cite{boyer,cb88,abe,cello,cb92}. Since the $\pi\pi$ system can be formed in both $I=0$ and $I=2$ final states, we have to treat the $\pi^+\pi^-$ and $\pi^0\pi^0$ channels simultaneously. Though there are now more than 2000 datapoints in the charged channel below 1.5 GeV, we only have 126 in the neutral channel, and we have to weight them more equally to ensure that the isospin components are reliably separable.

These world data can then be fitted adequately by a range of solutions~\cite{belle-mp}: a range, in which there remains a significant ambiguity in the relative amount of helicity zero $S$ and $D$ waves, particularly above 900 MeV.  
The acceptable solutions have a $\gamma\gamma$ width for the $f_0(980)$ (determined from the residue at the pole on the nearby unphysical sheet, being the only unambiguous measure) of between 96 and 540 eV, with 450 eV favoured: a significantly larger value than predicted by many current models~\cite{achasov,hanhart,sa,giacosa}. Of course, the experimental value includes the coupling of the $f_0(980)$ to its $\pi\pi$ and ${\overline K}K$ decay products and their final state interactions (the analogue of Fig.~3), not necessarily included in all the theoretical calculations.

The fits accurately follow the lower statistics data from Mark II~\cite{boyer} and Cello~\cite{cello} (see the detailed figures in Ref.~\cite{belle-mp}). However, they do not describe the Belle $\pi^+\pi^-$ data between 850 and 950 MeV, as seen in Fig.~1. This \lq\lq mis-fit'' is even more apparent in the angular distributions. In the charged pion channel, there is always a large $\mu^+\mu^-$ background. Though the Belle data have unprecedented statistics, the separation of the $\pi^+\pi^-$ signal is highly sensitive to the $\mu$-pair background. This may well be responsible for  the apparent distortion below 1~GeV in Fig.~1.

Since that analysis, Belle have more recently published results~\cite{belle00} (both integrated and differential cross-sections) on $\pi^0\pi^0$ production in 20 MeV bins, Fig.~1. Again these reveal the $f_0(980)$ as a small peak, rather than the shoulder seen in earlier much lower statistics data from Crystal Ball~\cite{cb88,cb92}. A new Amplitude Analysis has been started, which significantly changes the solution space, pushing the allowed amplitudes to those with a larger radiative width for the $f_0(980)$. However, we are not yet able to present the final solutions.

Though the Belle experiment 
represents an enormous stride in two photon statistics, there remains room for KLOE2 to make a significant contribution in each of the three energy regions displayed as bands in Fig.~1.
\begin{itemize}
\item[I.]  {\bf 850-1100 MeV}: accurate measurement of the $\pi^+\pi^-$ and $\pi^0\pi^0$ cross-sections (integrated and differential) are crucially still required, with clean $\mu\mu$ background separation. In addition, any information just above 1~GeV on $\overline{K}K$ production would provide an important constraint on the coupled channel Amplitude Analyses described above. Moreover $\pi^0\eta$ studies will complement the results to come from Belle.
\item[II.] {\bf 450-850 MeV}: this is the region where the $\sigma/f_0(600)$ pole lies. This is a region almost devoid of precision $\gamma\gamma$ data and so allows a range of interpretations~\cite{mp-prl,oller,menn}.
Given the importance of the $\sigma$ for our understanding of strong coupling QCD and the nature of the vacuum, it is crucial to  measure $\pi\pi$ production in this region in both charge modes~\cite{mpreviews}.
\item[III.] {\bf 280-450 MeV}: though this region is controlled by the Born amplitude with corrections computable by the first few orders of Chiral Perturbation Theory, it is the domain that anchors the partial wave analyses described here.
The Mark II experiment~\cite{boyer} is the only one that has made a special measurement of the normalised cross-section for the  $\pi^+\pi^-$ channel near threshold. As seen in Fig.~1, their data have very large error-bars.
\end{itemize}
The upgraded KLOE2 project can hopefully do better in terms of energy scan and precision in all three regions, and so add significantly to our understanding of low energy hadron physics.

\vspace{5mm}

The author acknowledges partial support of the EU-RTN Programme, Contract No. MRTN--CT-2006-035482, \lq\lq Flavianet'' for this work.

\end{document}